\documentstyle[aps,epsf]{revtex}
\def\be{\begin{equation}}
\def\ee{\end{equation}}
\def\bea{\begin{eqnarray}}
\def\eea{\end{eqnarray}}
\def\bml{\begin{mathletters}}
\def\eml{\end{mathletters}}
\def\sab{\sum_{\alpha<\beta}}
\def\sa{\sum_\alpha}
\def\lan{\langle}
\def\ran{\rangle}
\def\qab{q_{\alpha\beta}}
\def\pab{p_{\alpha\beta}}
\def\bJs{{\beta J^2}}
\def\bsJs{{\beta^2J^2}}
\def\D{{\cal D}}
\def\Tr{{\text Tr}}
\def\kT{{T}}
\def\pab{p_{\alpha\beta}}
\def\H{{\cal H}}
\def\Heff{{\cal H}_{\text{eff}}}
\def\Sa{\sum_a}
\def\Saa{\sum_{a\alpha}}

\def\qs{q_s}
\def\beps{{\beta\epsilon}}
\def\eps{{\epsilon}}
\def\bmu{{\beta\mu}}

\def\D{{\cal D}}
\def\bJsqz{{\beta J\sqrt{q}\,z}}
\draft
\title{Metastable states in the Blume-Emery-Griffiths spin glass model}
\author{Antonio de Candia}
\address{Dipartimento di Scienze Fisiche\\
INFM, Unit\`a di Napoli\\
Monte Sant'Angelo, I-80126 Napoli, Italy}
\begin{document}
\maketitle
\begin{abstract}
We study the Blume-Emery-Griffiths spin glass model in presence of
an attractive coupling between real replicas, and evaluate the effective
potential as a function of the density overlap. We find that 
there is a region, above the first order transition of the model,
where metastable states with a large density overlap exist. The line
where these metastable states appear should correspond to a purely dynamical
transition, with a breaking of ergodicity.
Differently from what happens in $p$-spin glasses, in
this model the dynamical transition would not be the precursor of a
1-step RSB transition, but (probably) of a full RSB transition.
\end{abstract}
\begin{figure}
\begin{center}
\mbox{\epsfysize=8cm\epsfbox{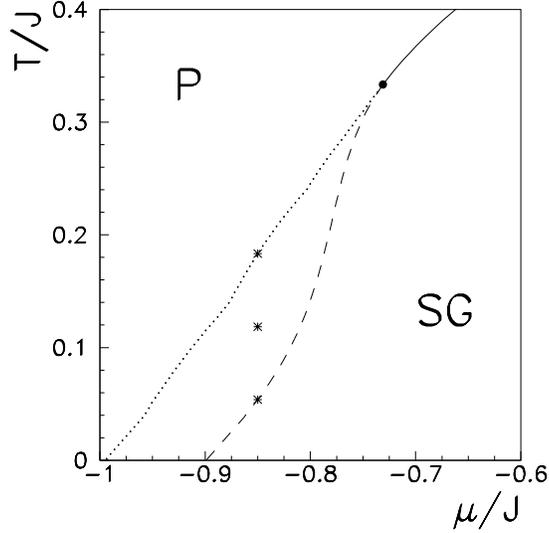}}
\end{center}
\caption{Phase diagram of the model, for $K/J=0$. The region P
is the paramagnetic phase, while the region SG is the spin glass phase.
The full curve
is the line of second order transitions, terminating at the tricritical point
(black dot). The dashed curve is the line of first order transitions,
and the dotted one is the line where non-paramagnetic solutions of
the saddle point equations first appear. The three asterisks are the points
where we have calculated the effective potential (see Fig. \ref{fig:vq}).}
\label{fig:beg}
\end{figure}

%
%
%
The Blume-Emery-Griffiths (BEG) spin glass model \cite{ref:mf1a,ref:mf1b}
can be seen as a generalization
of the Sherrington Kirkpatrick (SK) model, in which each site
carries, beside to the spin variable $S_i=\pm 1$, a lattice gas variable
$n_i=0,1$. Its Hamiltonian is given by
\be
\H=-\sum_{i<j}J_{ij}S_iS_jn_in_j-{K\over N}\sum_{i<j}n_in_j
-\mu\sum_{i}n_i
\label{eq:rowan}
\ee
where $J_{ij}$ are quenched Gaussian variables with zero mean and
variance $J^2/N$.
The case $K/J=0$ goes also under the name of Ghatak Sherrington model
\cite{ref:mf2}.
It has been found that in general,
fixed the value $K/J$,
there exists a tricritical point in the phase space. For 
higher temperature or chemical potential the model presents a second order
transition from a paramagnetic to a spin glass phase.
For lower temperature or chemical potential the transition becomes first
order, with a jump in the mean density $d=\lan n_i\ran$ and 
overlap $\qab=\lan S_i^\alpha n_i^\alpha S_i^\beta n_i^\beta\ran$
between two replicas $\alpha$ and $\beta$
(see Fig. \ref{fig:beg} for the case $K/J=0$).
Although the overlap parameter can become discontinuous, the transition seems
to be quite different from the transition of other discontinuous
spin glasses, like the $p$-spin models.
A first obvious difference is that the transition of the BEG spin glass 
is first order also in the Ehrenfest sense,
that is with a jump in the mean energy and entropy.
Moreover, while in $p$-spin models the first step of replica symmetry
breaking (RSB) turns out to be exact, in the BEG spin glass
the question of the order of RSB appears much more
subtle. The question was studied by many authors in the case
of the Ghatak Sherrington model \cite{ref:mf3,ref:mf4,ref:mf5}.
It was found that the replica symmetric solution is unstable
against RSB in the whole spin glass region, with
the appearance of complex eigenvalues in the stability matrix.
Recently \cite{ref:mf6}, it has been shown that this kind of
instability persists also in the 1-step RSB solution.
Therefore it seems reasonable that only a full replica symmetry breaking
may probably give the correct solution, as it
happens in the SK model.

In the last years, a version of the model called
frustrated Ising lattice gas (FILG), has been extensively studied
as a model of a structural glass \cite{ref:filg1a,%
ref:filg1b,ref:filg2,ref:filg4,ref:filg3a,ref:filg3b,ref:filg3c,ref:filg5}.
In this version of the model the disordered
interactions are of the type $J_{ij}=\pm J$, and $K=-J$.
Indeed, the FILG can be interpreted as a model
for a system of asymmetrical molecules,
with the variables $n_i$ representing the presence of a molecule on the
site $i$, and the variables $S_i$ representing the spatial orientation
of the molecule. Two neighbor particles feel a repulsion
of intensity $2J$ if their relative
orientation is ``wrong'', that is $J_{ij}S_iS_j<0$.
Numerical simulations on finite dimensional lattices
show that at low temperature the model develops
a two step relaxation \cite{ref:filg1a,ref:filg1b,ref:filg2,ref:filg4},
as observed in supercooled liquids and 
predicted by the mode coupling theory (MCT) of the glass transition
\cite{ref:gotze}. 

Given these links between the model and the glass transition in structural
liquids, it would be interesting to study (at least in mean field)
if the discontinuous transition of the model is
preceded at a higher temperature by a purely dynamical transition,
from an ergodic to a non-ergodic phase,
as it happens for $p$-spin glasses \cite{ref:kirks1,ref:kirks2}.
This would require
the solution of the dynamical equations, as done for the $p$-spin,
to see if they show a singularity, and of what kind, at some temperature
greater than the static one.
In this paper we address the problem from another point of view,
exploiting the effective potential theory, introduced some time ago
\cite{ref:effpot1a,ref:effpot1b,ref:effpot1c,ref:effpot2}
as a tool to identify the presence of metastable states, and 
a transition from an ergodic to a non-ergodic phase.
We will refer to the so-called ``annealed version'' of the method.
One considers two ``real'' replicas of the system (as opposed to the 
replicas generated by the replica trick), labeled 1 and 2, and defines
a ``degree of similarity'', that is an overlap $q$, between them. Then one
studies the system composed by the coupled replicas, with Hamiltonian
$\H=\H_1+\H_2-\epsilon q$. If $F(\epsilon)$ is the free energy of the
coupled system, then the Legendre transform
$V(q)=\max\limits_{\epsilon}[F(\epsilon)+\epsilon q]$ will represent the 
``effective potential'' as a function of the overlap $q$, that is
the work needed to keep the two replicas, in the configuration space,
at the distance $q$. What one finds in $p$-spin models
\cite{ref:effpot1a,ref:effpot1b,ref:effpot1c}
and in structural glasses in the HNC approximation
\cite{ref:effpot3a,ref:effpot3b},
and expects generally whenever metastable states are present in the system,
that is for temperatures between
the static and dynamical transition, is that
the potential $V(q)$ presents two minima.
A lower minimum, for low values of the overlap $q$,
corresponds to the ``unbounded state'', in which 
the two replicas are in different ``valleys'' of the configuration space.
A second higher minimum, for high values of $q$, corresponding
to the ``bounded state'', in which the two replicas are constrained to
stay in the same valley. As the free energy of each single system is
independent
from which valley the system is in, the gap $\Delta V$ between the two minima
can be interpreted as being equal to $T\Sigma$, where $\Sigma$ is the 
logarithm of the number of valleys, and is called complexity
or configurational entropy.

We apply the effective potential theory to the model defined by 
(\ref{eq:rowan}) with $K/J=0$. The phase space is depicted in 
Fig. \ref{fig:beg}, where the continuous curve represents the line of
second order transition, the black dot the tricritical point, and 
the dashed curve the line of first order transitions. We now introduce
two real replicas,
coupled in the density overlap $q_d={1\over N}\sum_i n_i^1n_i^2$.
The Hamiltonian of the system is
\be
\H=-\sum_{a=1,2}\left(\sum_{i<j}J_{ij}S_i^aS_j^an_i^an_j^a
+\mu\sum_{i}n_i^a\right)
-\eps\sum_{i}n_i^1n_i^2
\ee
where $a=1,2$ is the label of the real replica.
By the usual replica trick, we can write the free energy of the system as
\be
F={\bJs\over 2}(d^2+\qs^2)
+{\bJs\over n}\sab(\qab^2+\pab^2)-{\kT\over n}\log\Tr\exp(-\Heff)
\ee
where $d=\lan n^{a\alpha}\ran$ is the mean density,
$\qs=\lan S^{1\alpha}S^{2\alpha}n^{1\alpha}n^{2\alpha}\ran$
is the spin overlap between two directly coupled replicas,
$\qab=\lan S^{a\alpha}S^{a\beta}n^{a\alpha}n^{a\beta}\ran$
and
$\pab=\lan S^{1\alpha}S^{2\beta}n^{1\alpha}n^{2\beta}\ran$
are respectively the overlap matrices between replicas with the same
and with different ``real label'',
and $\Heff$ is
the effective single site Hamiltonian
\bea
\Heff=&&-\left({\bsJs\over 2}d+\bmu\right)
\Saa n^{a\alpha}
-\bsJs\sab\qab\Sa S^{a\alpha}S^{a\beta}n^{a\alpha}n^{a\beta}
\nonumber\\
&&\mbox{}-2\bsJs\sab\pab S^{1\alpha}S^{2\beta}n^{1\alpha}n^{2\beta}
-\bsJs\qs%
\sa S^{1\alpha}S^{2\alpha}n^{1\alpha}n^{2\alpha}
-\beps\sa n^{1\alpha}n^{2\alpha}
\eea
In the replica symmetric approximation $\qab=q$ and $\pab=p$.
We make the further approximation $q=p$, and obtain for the free energy
\bea
F=&&-2\kT\log 2+{\bJs\over 2}(d^2+\qs^2-2q^2)
\nonumber\\
&&\mbox{}-\kT\int\!\D{z}\,\log\left\{%
1+2e^\Xi\cosh(\bJsqz)
+e^{2\Xi+\beps}\left[e^\Omega\cosh^2(\bJsqz)
-\sinh\Omega\right]\right\}
\label{eq:free}
\eea
where $\Xi={\bsJs\over 2}(d-q)+\bmu$,
$\Omega=\bsJs(\qs-q)$,
and $\D{z}={dz\over\sqrt{2\pi}}\,e^{-z^2/2}$.

The physical states of the system will be given by the saddle points
of the free energy, which are given by the equations
\bml
\label{eq:saddle}
\bea
d&=&\int\!\D{z}\>\>{e^\Xi\cosh(\bJsqz)
+e^{2\Xi+\beps}\left[e^\Omega\cosh^2(\bJsqz)
-\sinh\Omega\right]\over
1+2e^\Xi\cosh(\bJsqz)
+e^{2\Xi+\beps}\left[e^\Omega\cosh^2(\bJsqz)
-\sinh\Omega\right]}\\
q_s&=&\int\!\D{z}\>\>{e^{2\Xi+\beps}\left[e^\Omega\cosh^2(\bJsqz)
-\cosh\Omega\right]\over
1+2e^\Xi\cosh(\bJsqz)
+e^{2\Xi+\beps}\left[e^\Omega\cosh^2(\bJsqz)
-\sinh\Omega\right]}\\
q&=&\int\!\D{z}\,\left({e^\Xi\sinh(\bJsqz)
\left[1+e^{\Xi+\beps+\Omega}\cosh(\bJsqz)\right]\over
1+2e^\Xi\cosh(\bJsqz)
+e^{2\Xi+\beps}\left[e^\Omega\cosh^2(\bJsqz)
-\sinh\Omega\right]}\right)^2
\eea
\eml
Notice that the saddle points are neither maxima nor minima of the free energy,
because $F$ has to be minimized with respect to $d$ and $q_s$, and
maximized with respect to $q$.
If $\epsilon=0$
then $q_s=q$, $\Omega=0$,
and one obtains the saddle point equations of the uncoupled system.
Differentiating the free energy $F$ with respect to $\epsilon$, we can
evaluate the density overlap $q_d=\lan n^{1\alpha}n^{2\alpha}\ran$,
\be
q_d=-{\partial F\over\partial\epsilon}=
\int\!\D{z}\>\>{e^{2\Xi+\beps}\left[e^\Omega\cosh^2(\bJsqz)
-\sinh\Omega\right]\over
1+2e^\Xi\cosh(\bJsqz)
+e^{2\Xi+\beps}\left[e^\Omega\cosh^2(\bJsqz)
-\sinh\Omega\right]}
\ee
In the above expression we can put $\epsilon=\Omega=0$, and find the
expression of the density overlap for a system without coupling, which
cannot be obtained directly from the free energy of the uncoupled system.

\begin{figure}
\begin{center}
\mbox{\epsfysize=8cm\epsfbox{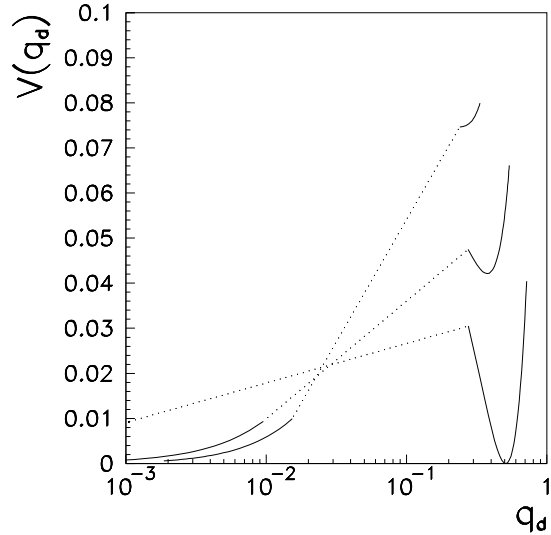}}
\end{center}
\caption{Effective potential as a function of the density overlap $q_d$,
for chemical potential $\mu/J=-0.85$ and temperatures
$T/J=0.183$, 0.119 and 0.0538 (points marked with an asterisk
in Fig. \ref{fig:beg}). The potentials for different temperatures are shifted
so that the paramagnetic minimum is at $V=0$. The dotted lines join
the potentials at the same temperature, and are only a guide to the eye.}
\label{fig:vq}
\end{figure}

Now we want to compute the effective potential as a function of
the density overlap $q_d$. The procedure to do so was previously
sketched, and consists in evaluating $V(q_d)=\max\limits_{\epsilon}
[F(\epsilon)+\epsilon q_d]$. We evaluate $V(q_d)$ in three points of the phase
space (marked by asterisks in Fig. \ref{fig:beg}). One point at the 
edge of the region in which spin glass solutions of the saddle point
equations exist, one on the line of first order
transitions, and one in between. The result is shown in Fig. \ref{fig:vq}.
As soon as one enters in the region 
where spin glass solutions exist, a secondary minimum in the 
effective potential appears, signaling the existence of metastable states.
We expect therefore that the dotted line of Fig. \ref{fig:beg}
will represent a line of purely dynamical transitions, below which
ergodicity is broken and the relaxation function
$\lan n_i(0)n_i(t)\ran$ will not decay to its equilibrium value,
which in the paramagnetic phase is the density squared, but will 
stick at the value given by the secondary minimum of the potential.

The presence of metastable states, and of a dynamical transition 
at a temperature higher than the static transition, in a model in which
the low temperature spin glass phase is
characterized by a full RSB, seems rather peculiar.
It is reasonable to think that the usual picture,
valid for $p$-spin models, according to which at the dynamical
transition the phase space is splitted
into an extensive number of valleys having all the same
``distance'', that is the same mutual overlap,
is modified in this case. Here in fact
the spin glass phase is characterized by a complex ultrametric structure,
typical of the SK model.
In particular one may ask if in this case the gap $\Delta V$
in the effective potential can still be
interpreted as a configurational entropy, that is
the logarithm of the number of valleys in the configuration space.
Finally, it would be interesting to understand how and
to what extent this model can tell us
something about real systems, that is structural glasses.

I thank A.~Coniglio, M.~Nicodemi, M.~Sellitto and J.J.~Arenzon for 
many interesting discussions.
This work was partially supported by the European TMR Network-Fractals
(Contract No.\ FMRXCT980\-183), MURST-PRIN-2000 and INFMPRA(HOP).

%
%
%
%

%
%
%
%

\end{document}